# Light-enhanced van der Waals force microscopy

Yu-Xiao Han, Benfeng Bai*, Jian-Yu Zhang, Jia-Tai Huang, Peng-Yi Feng, Hong-Bo Sun*

*State Key Laboratory of Precision Measurement Technology and Instruments, Department of Precision Instrument, Tsinghua University, Beijing 100084, China*

*Corresponding author: baibenfeng@tsinghua.edu.cn; hbsun@tsinghua.edu.cn.*

**Abstract**

Atomic force microscope (AFM) generally works on the basis of manipulating absolute magnitude of van der Waals (vdW) force between the tip and specimen. The force is, however, less sensitive to alternation of atom species than to tip-sample separations, resulting in the difficulty of compositional identification, even under multi-modal strategies and other AFM variations. Here, we report a phenomenon of light enhancement of van der Waals force (LvF), and the enhancement factor is found specific to materials. The force difference prior and after illumination, instead of the tip-specimen force itself, is employed for discriminating heterogeneous phases. The corresponding LvF microscopy (LvFM) demonstrates not only a ultra-high compositional resolution represented by 20 dB enhancement factor and 150 times of the detection limit, but also a sub-10 nm lateral spatial resolution much smaller than the tip size of 20 nm. The simplicity of the opto-thermal mechanism, minuteness of excitation light power and wide availability of boosting lasers at various wavelengths imply broad applications of LvFM on nano-materials characterization, particularly on two-dimensional semiconductors that are promising as new generation of chip materials.

Since the early speculation by Binnig, Quate and Gerber in 1986, atomic force microscope (AFM)[1] has been established as a unique nano-characterization tool with spatial resolution more than 1000 times better than the optical diffraction limit[2, 3, 4]. It utilizes mainly van der Waals (vdW) force[5], $F_v$, between the sharp tip apex of a cantilever and the specimen surface for force measurement and morphology mapping. Although vdW force is the weakest among the long-range forces, it is sensitive to distance between atoms, $r$, by law of $F_v \sim r^{-7}$. This leads to, on one hand, a vertical resolution high enough to recognize single-atom steps[6, 7], and on the other hand, the difficulty in compositional identification at ambient temperature and atmosphere, for example, heterogeneous phases embedded in mono- or few-layer two-dimensional (2D) semiconductors[8, 9], due to the morphological perturbation.

For the purpose of composition recognition, alternative strategies have been developed[10, 11]. For example, AFM-based infrared spectroscopy (AFM-IR)[12] resorts to molecular resonant absorption to locally swell sample surface, and scanning near-field optical microscope (SNOM)[13, 14] monitors fluorescence or Raman scattering. Both reach the end to some extent but lose the simplicity or versatility of AFM, for the price of needing multiple wavelengths to match particular infrared absorption peaks of different materials (AFM-IR) or requiring spectrometers for wavelength separation (SNOM). Multi-modal AFM goes further along this line. Rodríguez and García[15] found that by simultaneously exciting fundamental and high-order vibration modes of a cantilever, material compositions were distinguishable from the phase imaging of the latter. Despite the success of bi-modal AFM in KBr (001)[16], flexible crystals [Cu (acac)$_2$][17], human teeth[18], metal organic framework[19, 20], and single IgM antibody protein[21] imaging, it is suffering from the sensitive dependence on the intrinsic property discrepancy of compositional materials for sufficient phase contrast, and is only applicable to limited types of materials.

Here, we report a novel phenomenon, i.e., light-enhaced vdW force (LvF), which leads to an effective solution to the challenge of compositional identification. The enhancement is attributed to asymmetric force variation due to the minus 7$^{th}$ power dependence on tip-sample distance[22] when sample surface atoms approach nearer and depart further from the AFM tip with greater amplitude in thermal motion enhanced by the optothermal effect. The net force increment is found sensitively dependent on materials, laying the physical foundation for the novel concept of light-enhanced van der Waals force microscopy (LvFM). The effectiveness of the LvFM lies in involving chemical bonding forces[23] in addition to vdW force, and magnifying the difference of intrinsic material characteristics by light boosting. A compositional resolution 150 times higher than the detection limit is thus achieved. A comparison of LvFM with other methods is discussed in the supporting information, SI-1.

**Two types of atomic forces**

Both vdW force and chemical bonding force (Fig. 1a) are widely known atomic forces to bind atoms or molecules into condensed phase[24]. The former, often appearing as an inter-molecular force, governs the stability of colloids and dominates energies of surfaces and interfaces[25]; while the latter, for the sake of brevity, bonding force, creates mechanical resist against compression, extension, bending and distortion of solid materials[26]. vdW force is produced due to transient separations of positive and negative centers in neutrally charged atoms, molecules or particles, which pairwise induce dipoles spontaneously in a neighboring object, acting to and reacting on it[5]. The second-order-perturbation nature of the transient dipole pair interaction results in sharp, minus 6th-power in potential, and minus 7th-power in force, dependence on separation (Fig. 1b). The bonding force, on the other hand, arises in principle from strong Coulombic interactions among positively charged atomic cores and redistributed negative valence electron clouds[27]. It presents either attractive or repulsive depending on whether a solid sample is pulled long or compressed. The repulsive bonding force also appears when a tip approaches a sample surface within the lattice constant approximity, with a minus 13th-power dependence on distance. The overall tip-sample forces are presented by Eq. 1 as follows[22],

$$f_{\mathrm{LJ}} = 12\frac{\varepsilon}{r_0}\left(\frac{r_0^7}{r^7} - \frac{r_0^{13}}{r^{13}}\right), \tag{1}$$

where $\varepsilon$, $r_0$, and $r$ are the bond energy of Lennard-Jones potential, the inter-atomic distance at potential equilibrium, and the practical inter-atomic distance, respectively. The first term at the right side of Eq. 1 is an attractive vdW force, $F_V$, and the second is a repulsive bonding force, $F_B$. The latter determines the amplitude of an AFM cantilever in tapping mode[28] (Fig. 1b) when an appropriate tip-sample distance is chosen, as is utilized as a measure of vdW force. More information is given in SI-2.

It is worthy to mention that vdW force occurring between identical and alien atoms or molecules is generally less than the attractive bonding force by orders[27], and their huge difference in magnitude explains why they are seldom mutually adopted to characterize materials and their interactions. The dilemma of composition identification in AFM arises from the fact that only vdW interactions between the tip and specimen are employed, which reflects partial characters of matter to be detected. The bonding force, $F_B$, indiscriminately imposed on surface atoms by their neighborhood beneath, is playing an even more important role in materials performance and their response to external stimuli. If it is boosted and appropriately included in characterization, together with vdW force, new way may be paved towards effective compositional identification. Here, the mismatch has actually been overcome by the light boosting bonding force. It is the bonding force change, instead of the force itself, that becomes comparable to vdW force, and further leads to occurrence of LvF.

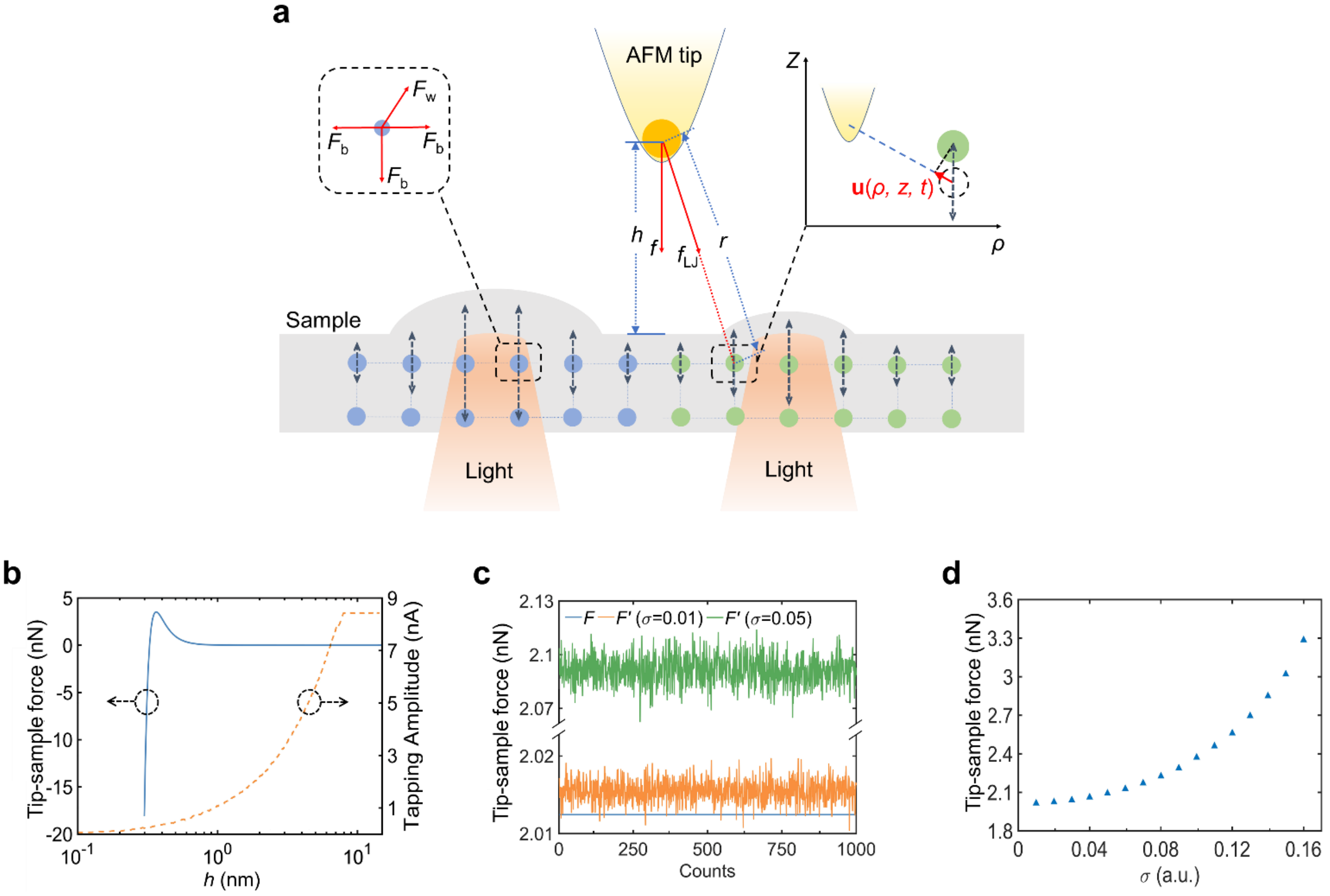

**Fig. 1|** Physical original of LvF. **a.** Schematic of vdW force interactions between an AFM probe tip and a sample surface when their distance is small enough. The force is enhanced by light excitation due to the strengthened atomic thermal motion, of which the normal component of their amplitude is denoted by the vertical arrows. The enhancement factor is sensitively dependent on materials (blue and green circles) under identical illumination. **b.** Calculated total tip-sample force $F$ versus tip-sample distance $h$ (blue), and practical tapping amplitude of an AFM probe versus the tip-sample distance (orange). In calculation, $\varepsilon = 1\times10^{-19}$ J, $r_0 = 0.3$ nm, and the tip radius $R = 10$ nm. **c.** Monte Carlo simulations of optothermally enhanced total force $F'$ when $\sigma = 0.01$ and $\sigma = 0.05$. The $F$ curve is provided for comparison. The distance between tip and sample is fixed as 1 nm. **d.** $F'$ versus the intensity of atomic thermal motions described by $\sigma$. $F'$ is the average value of 1000 times Monte Carlo simulation at different $\sigma$.

**vdW force enhanced by light**

Light boosting brings about local vdW force enhancement when a laser is focused beneath the tip on a sample, as is manifested analytically and numerically as follows. The vdW force between the AFM tip and a sample requires integrating the dipole interactions over the tip profile (hemisphere, for example) and the infinitely semi-space,

$$F = \int n_1 dV_1 (\int n_2 dV_2 \cdot f_{\mathrm{LJ}}) = \int n_1 dV_1 \cdot f \,, \tag{2}$$

where $n$ and $V$ represent the dipole (atom) density and integral volume of the probe tip (subscript 1) and the sample (subscript 2), respectively; $f$ ($f'$) and $F$ ($F'$) are therefore the force between a certain atom in the tip and the whole specimen and the entire tip-sample force without (with) light illumination, respectively, as follows,

$$f' = \iiint_{\Omega} n_2 \mathbf{f}_{\mathbf{LJ}}(r + u(\rho, z, t)) \cdot \mathbf{n} dV \triangleq \frac{\pi n_2 \varepsilon r_0^6}{12(h - \Delta h)^4} \tag{3}$$

where $\boldsymbol{f}_{\mathbf{LJ}}(r+u(\rho,z,t))$, $\boldsymbol{n}$, $h$, $\Delta h$ are the vector of Lennard-Jones force, unit vector of sample surface normal, tip-sample distance, and overall decrement of the tip-sample distance due to light illumination, respectively. $\Omega$ is the entire sample volume. Since $\Delta h$ is always larger than zero when a sample is photo-thermally excited, Eq. 3 indicates clearly the existence of LvF, i.e., $\Delta F = F' - F$ . The detailed calculation of force is given in SI-3. The underlying physics lies in the fact that atoms are pushed nearer to the tip in one half-period, and further in another half when thermal vibrations are intensified. While the later leads to a force decrease, to a smaller but still positive value, the force increase in the former over-compensates the loss due to the asymmetric $7^{\text{th}}$-power dependence.

Based on Eqs.1-3, the generation of LvF can be validated by Monte Carlo simulations[29]. The random amplitude variation of atomic thermal motion of $u(\rho, z, t)$ is assumed to follow Cauchy distribution, whose mathematical expectation is 0, reflecting the fact that the thermal motion is always around its mass center. When $u(\rho,z,t)$ is taken as 0.01, 1000-time dynamic simulation shows (Fig. 1c) $\Delta F > 0$ for 982 times $\Delta F < 0$ for only 18 times. This confirms that light irradiation does enhance the tip-sample vdW force. As numerical examples, time-averaging $F'$ is increased by 0.15% and 3.9% over $F \approx 2$ nN when $\sigma = 0.01$ and 0.05, respectively (Fig. 1d), and the enhancement factor is exponentially dependent on the incident light power. It is also noticeable that $u(\rho, z, t)$ and $\Delta h$ are specific to materials (Fig.1d), and it is expected that they are positively correlated to the linear thermal expansion coefficient of a solid, as well as their light absorptivity. This fact is of fundamental importance for compositional identification, as detailed later. Further discussion is given in SI-4,5.

## Picking up of LvF signals

Instrumentally, bare sample surface morphology and LvF are synchronously read via a dual-modal cantilever with fundamental and second order frequencies of $f_1$ and $f_2$, facilitating in-site and real-time recording images. For this purpose, we established a home-built system based on an AFM platform and equip it with an inverted optical microscope (Figs. 2a,b). Its probe installed on a dual-modal cantilever works in tapping mode, driven by a dither piezo at frequency $f_1$. A laser beam with modulation frequency $f_m$ is focused onto a sample and illuminates it from the bottom. Coupling of forces between $f_1$, originating from the tapping, and $f_m$, from LvF results in beating frequencies of $f_2 = |f_m \pm f_1|$ (Figs. 2c), at which the maximum of transiently enhanced vdW force signal appears.

Experimentally, we use a silicon probe with $f_1$ = 235.7 kHz, quality factor $Q_1$ = 572 and $f_2$ = 1471.8 kHz, $Q_2$ = 756 (Fig. 2c). The set point of the proportional integral derivative controller, used to maintain a constant tip-sample distance during scan, is selected as 60% of the free oscillation amplitude (~10 nm). This ensures that the probe enters the vdW force interaction zone in one oscillation cycle. The boosting laser is modulated at $f_m = f_2 - f_1$ = 1236.1 kHz. A monitoring laser beam is reflected from the back of the AFM cantilever and detected by a position sensitive detector (PSD). By this means, the cantilever oscillation is monitored and amplified by the optical lever as an electric signal, which is then split and demodulated by two lock-in amplifiers (Figs. 2a,b). One of them takes $f_1$ as the reference to measure the stationary surface morphology, while the other refers to $f_2$ provided by an electronic mixer, mixing $f_m$ and $f_1$ to detect the transient LvF topography. More information is given in SI-6.

Since the amplitude and phase of the transiently enhanced force signal are demodulated at $f_2$ while the tapping frequency is fixed at $f_1$, we scan the laser modulation frequency $f_m$. Signal maximum appears at 1236 kHz, exactly the value of $f_m = f_2 - f_1$, exhibiting the effectiveness of the dual-channel signal pick up (Fig. 2d). A detailed discussion is provided in SI-7. Furthermore, a high smoothness borosilicate glass plate is chosen as a sample to experimentally certify the LvF existence. Its surface roughness, measured with a commercial AFM, is at sub-nanometer level (Fig. 2e), identical to that from the $f_1$ channel of LvFM (Fig. 2f). Astonishingly, a symmetric cone-like feature appears upon the boosting laser turning on (Fig. 2f). Its geometrical shape is basically a copy of the Gaussian laser beam profile, showing its force feature instead of any actual physical protrusion since Fig. 2e, recorded simultaneously with $f_1$, exhibits negligible surface roughness.

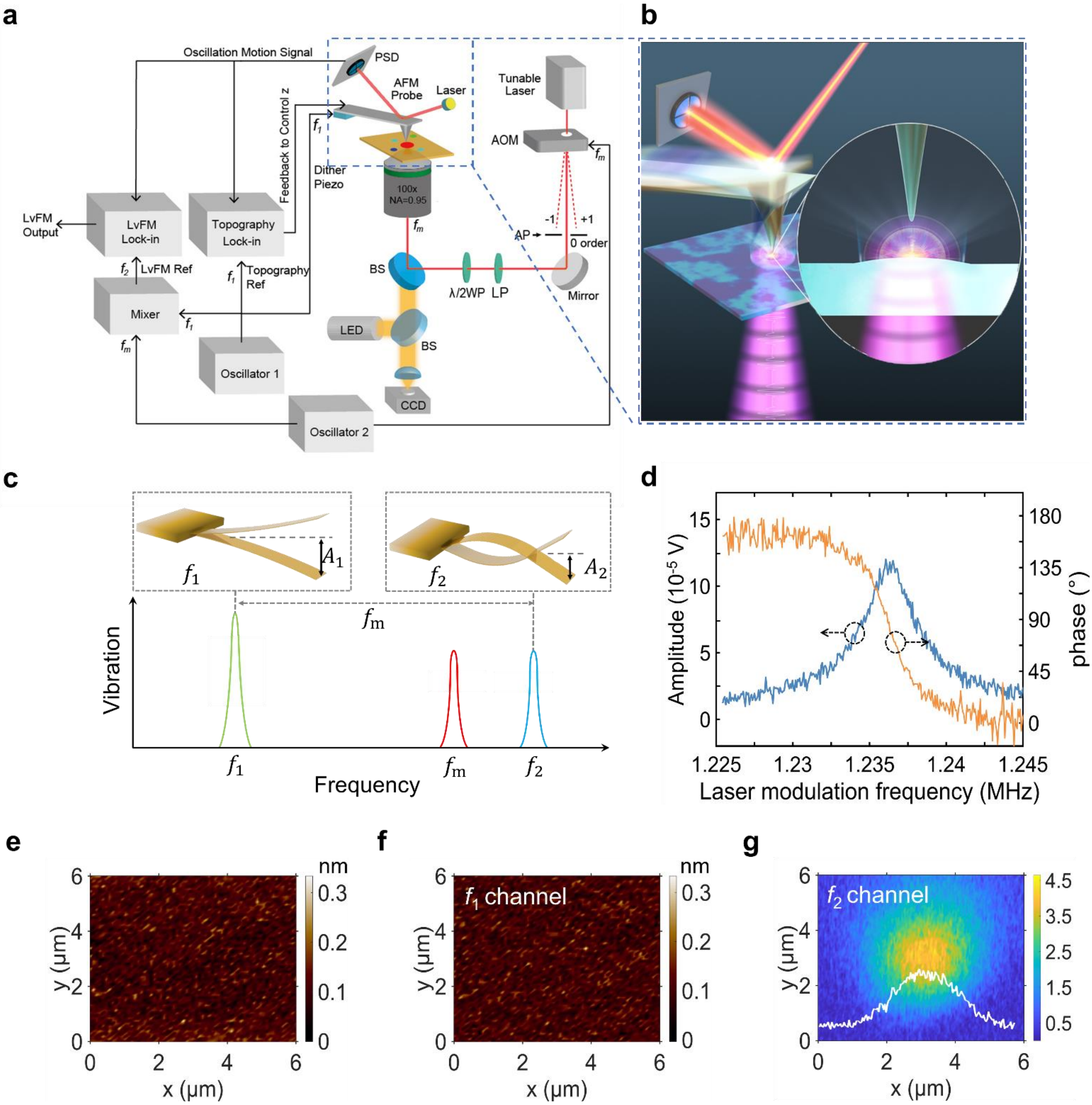

**Fig.2| Instrumental realization of LvFM. a,** LvFM setup. AOM: acousto-optic modulator; PSD: position sensitive detector; AP: aperture; WP: wave plate; LP: linear polarizer; BS: beam splitter; LED: light emitting diode; CCD: charge coupled device. $f_1$ is the tapping frequency of the probe. $f_m$ is the modulation frequency of the illuminating laser. **b,** Schematic of the LvFM working principle. The local LvF is detected by demodulating the AFM probe oscillation at $f_2$ that is amplified by an optical lever. **c,** Illustration of an AFM cantilever with two mechanical resonance eigenmodes with frequencies $f_1$ and $f_2$ and resonance amplitudes $A_1$ and $A_2$. The modulation frequency $f_m$ is chosen such that $f_2 = |f_m \pm f_1|$. **d,** Measured light-boosting LvF response curves of a probe on a borosilicate glass, which show the amplitude and phase of the force signal versus the laser modulation frequency $f_m$ when the tapping frequency is fixed at $f_1$. **e,** The height image of borosilicate glass obtained by a commercial AFM. **f**, **g,** Height image and LvFM image of a smooth borosilicate glass with boosting laser of 1.2 mW at 633 nm.

**LvFM spatial resolution and compositional resolution**

LvFM possesses unique compositional resolution besides its amorphous resolution identical to general AFM (Figs. 2e,f). The ability is a direct result of LvF associated with and specific to atoms of varied bonding status, either from different atomic species or from identical atoms of different phases. The lateral resolution is firstly improved by the boosting laser focal spot to the level of optical diffraction limit, and is finally defined by the tip-sample interactions. The limit is imposed by finitely-small tip, which is attracted by atoms from two sides when it scans across their boundary (Fig. 3a). It is estimated smaller than the 10-nm width of the slope between the 2H and 1T' phases contained in a flat 10-nm-thick $MoTe_2$ polycrystal film (Fig. 3b), which is unexpectedly smaller than the 20 nm tip apex diameter. In contrast, the lateral resolution of AFM is restrained by deconvolution of tip shape and surface profile, and that in AFM-IR is exaggerated by materials thermal diffusion length, while LvFM demonstrates for the first time a lateral resolution much smaller than the tip size. The theoretical limit calculated according to Eqs. 1-3 is roughly 4 nm (SI-8), meaning more potential to improve solely by sharpening the probe.

The longitudinal resolution is meaningless for compositional identification in a flat surface, instead it appears as composition resolution, defined as LvF ratio by $\delta_1$:$\delta_2$ ($\delta_1 > \delta_2$). Measured from repetitive laser on-off, a 20 dB signal-to-noise ratio (SNR), also meaning the enhancement factor, is attained from our system even if the boosting light power is only 1.3 mW (Fig. 3c), a level much lower than those in usual fluorescence or Raman spectroscopies. It sets a detection limit, DL = $(\delta_1-\delta_2)/\delta_1$ ~ 1% in composition recognition. Such a low excitation level LvF and the composition resolution are the functions of bonding force, amount of boosted bonds, light intensity, and material absorptivity. It is worthy to mention LvF is boosted by superimposing illumination of directly incident and scattered lights, impling that higher scattering intensity leads to better composition resolution for a given initial incident light intensity.

These scenario can be approved by 6 combinations of 3 sets of parameters, 2 materials (2H and 1T' phases of $MoTe_2$), 2 thickness (10 nm and 5 nm), and 2 probes (Au and Si). For example, (i) $\delta_{2H}$:$\delta_{1T'}$ = 2.58 under 10 nm sample thickness, Au probe, 1.3 mW laser power, and $(\delta_{2H}-\delta_{1T'})/\delta_{1T'}/DL$ = 158, a resolving power more than 150 times stronger than the detection limit; (ii) $\delta_{10nm}$ :$\delta_{5nm}$ = 2.56 (2H, Au, 1.3 mW); (iii) $\delta_{Au}$ :$\delta_{Si}$ = 1.45 (2H, 10 nm, 1.3 mW); and (iv) the LvF dependences on the laser power slopes in unit of $10^{-5}$ V/mW are 1.59 (10 nm, 2H, Au), 0.66 (10 nm, 1T', Au), 0.6 (5 nm, 2H, Au), 0.36 (5 nm, 1T', Au), 1.08 (10 nm, 2H, Si), and 0.35 (10 nm, 1T', Si), respectively. It means that 2H phase is "softer" than 1T', large thickness includes more bonds boosted, and Au probe scatters light more strongly than Si probe. It is worthy to mention that signals in all circumstances are sufficiently robust (Fig. 3d), allowing long-time mapping of 2D images.

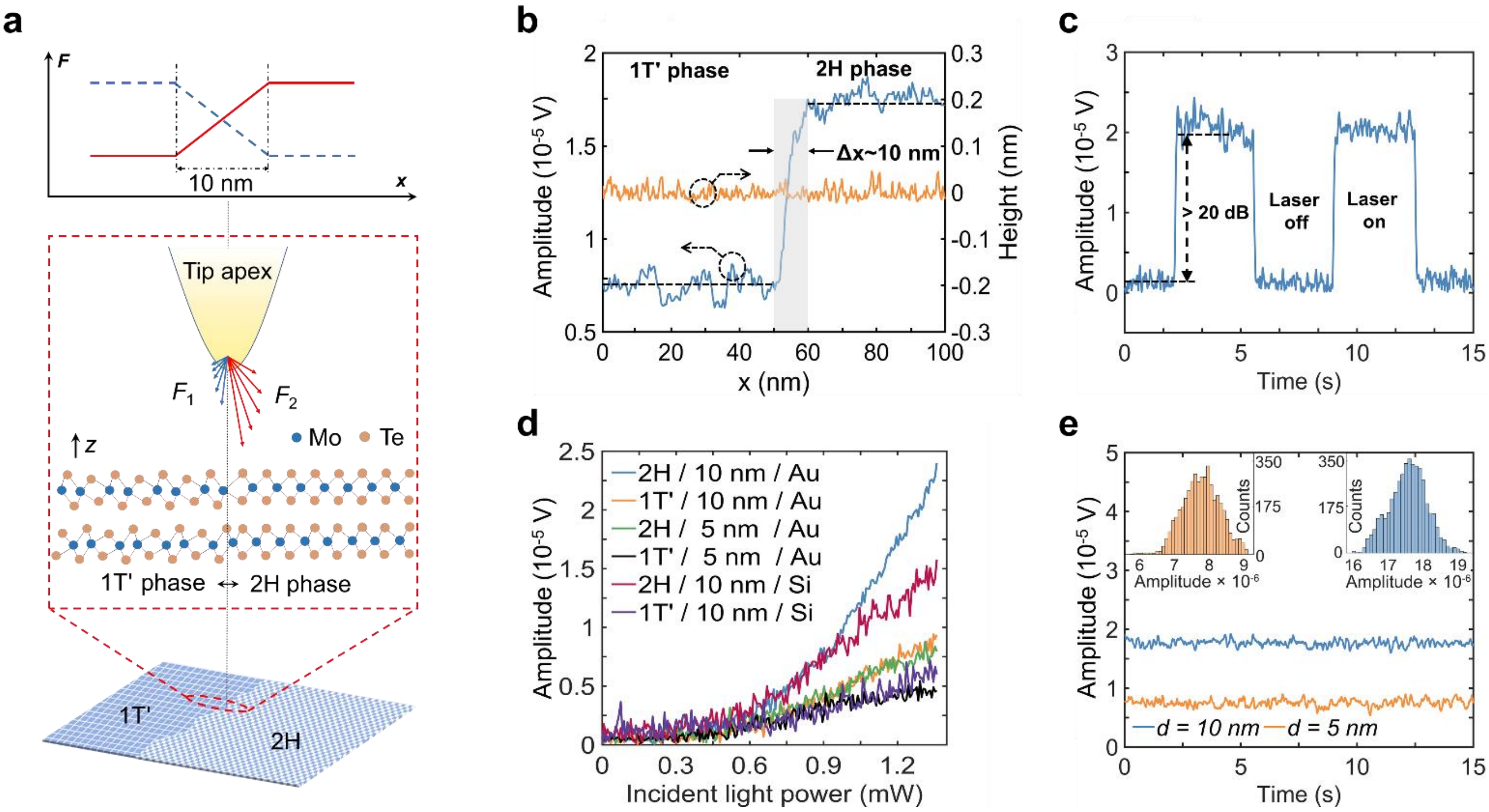

**Fig. 3 | Compositional resolution of LvFM. a.** Schematic illustration of the two crystalline phases 2H and 1T' in a $MoTe_2$ film. $F_1$ and $F_2$ represent the forces exerted on the tip from the 1T' and 2H phases when the tip crosses their boundary. The force may increase (solid line, the top inset) or decrease (broken line) dependent on materials. **b.** Measured LvF amplitude profile when the probe scans across a boundary of the 2H and 1T' phases with the step of 0.5 nm. The shaded area indicates an estimate of the 10 nm lateral resolution. The used tip radius is about 10 nm. **c.** LvFM amplitude response when the laser switches on or off, where the ratio of the high to low voltages is defined as the compositional resolution. The tip and the sample are both fixed at the lateral positions during the test. **d.** Measured LvF amplitudes versus the power of incident light for 6 experiments with different combinations of $MoTe_2$ crystalline phases (2H or 1T'), film thicknesses (10 nm or 5 nm), and probe tip materials (Au or Si). **e,** Temporal stability of the measured LvF signal amplitudes for samples with 5 nm and 10 nm thicknesses. The fluctuations of the LvF from the 10 nm and 5 nm thick $MoTe_2$ films do not exceed 8% and 13%, respectively. The insets show the histograms of the measured LvF amplitude data.

**Compositional identification of 2D semiconductors by LvFM**

LvFM satisfies urgent needs of composition and nano-defects characterization of nanomaterials, particularly 2D semiconductors[30, 31], for example, the occasion of inserting a nanoscale metal-phase buffer layer between the channel and electrode in manufacture of field effect tubes[32]. Abundant nano-defects including nanobubbles, stacks, vacancies, grain boundaries, impurities and adsorptions are associated with material preparation and device fabrication[33]. They are not completely characterizable by currently available technologies. X-ray diffraction is good at determination of crystalline structures but lack of spatial resolution; Electron microscopes possess the highest morphological resolution better than 1 nm, but destructive sample preparation is needed for composition discrimination[34]; AFM-IR is suffering from relatively low resolution and strict requirement of wavelength provision[35].

In contrast, boosting bonding force to enhance vdW force in LvFMrequires relatively weak light excitation that are possibly provided by lasers of almost any wavelength. Light absorption mechanisms ranging from inter-band transitions, impurity absorption, molecular vibrations and rotations all work for LvFM. We choose a 650 nm visible laser for compositional identification from an atomically smooth $MoTe_2$ (Figs. 4a,b), whose preparation and characterization are given in SI-9. The $f_1$ channel gives rise to 2D stationary morphological images (Fig. 4a) of the same quality as that obtained by commercial AFM, while the $f_2$ channel demonstrates clear 2H and 1T' phase distributions by 300 × 300 pixels in an area of 20 μm × 20 μm within 14 min (Fig. 4b). The identification is further confirmed by Raman peaks at 235 $cm^{-1}$, attributed to in-plane $E_{2g}$ mode of 2H phase, and 124 $cm^{-1}$ ($A_g$) and 163 $cm^{-1}$ ($B_g$) associated with 1T' phase (Fig. 4c), as well as by its confocal mapping (Fig. 4d). Tuning the laser wavelength from 500 nm to 900 nm results in similar level of compositional resolution (see Figs. 4e, f, h and SI-10), for example, $\delta_{2H}$ : $\delta_{1T'}$ = 1.08 (520 nm), 1.33 (633 nm), and 1.40 (730 nm), showing the wide spectroscopic applicability.

Compositional identification from a non-flat mixture is even challenging since the LvF signal tends to be deluged into amorphous height fluctuation in conventional AFM and in $f_1$ mode of LvFM. This is because the LvF contribution to the stationary morphology is negligibly small according to the previous Monte Carlo simulations. A medium light boosting with $u = 0.01$ and $\sigma = 0.05$, meaning bond extension of the same amount, approximately, 0.001 nm, is at least 2-order less than the height fluctuation magnitude of several to tens of nanometers. While tens of nanometer vertical height fluctuation is attained from the $f_1$ channel images (Fig. 4g), h-BN and $WS_2$ multilayer nanoflakes are well distinguished from their mixture prepared on a borosilicate glass substrate from the $f_2$ channel (Fig. 4i). Further characterization of the $WS_2$/h-BN heterogenous materials is given in SI-11.

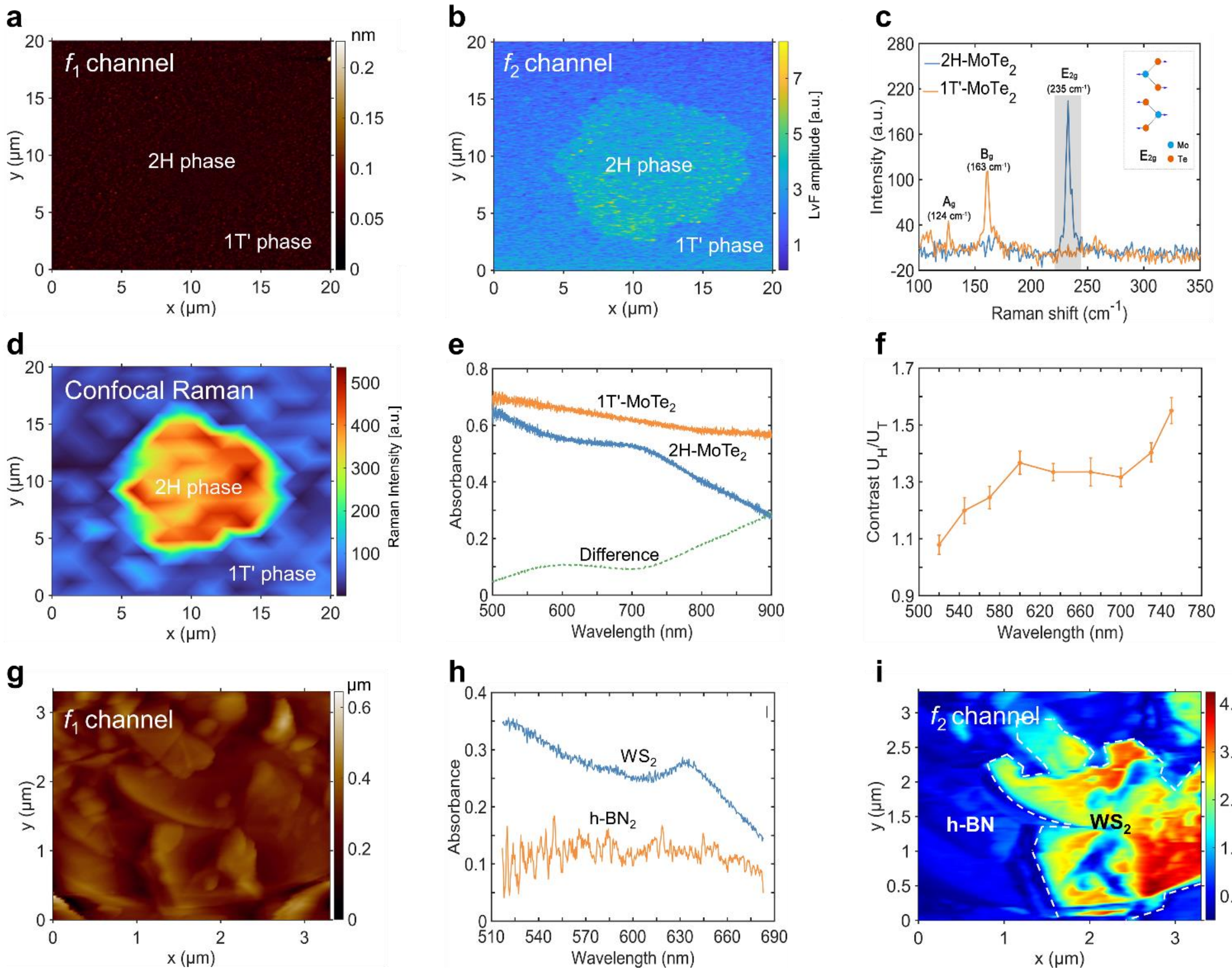

**Fig. 4 | Discrimination of the heterophases and heterogeneous materials. a.** Measured topography of a 10 nm thick $MoTe_2$ film with two crystalline phases using LvFM $f_1$ channel, indicating the surface flatness. **b.** $f_2$ channel signal mapping under 650 nm laser excitation with approximately 1.2 mW, where the 2H phase and 1T' phase regions are clearly discriminated. **c.** Raman spectra from different phases, where the fingerprint Raman peaks distinguish the 2H and 1T' phases. **d.** Confocal Raman mapping image obtained with Raman shift from 225 $cm^{-1}$ to 245 $cm^{-1}$, indicated in the shaded area in **c**. **e.** Measured absorbance spectra of the 2H phase (blue curve) and 1T' phase (orange curve) of the $MoTe_2$ film, whose difference is indicated by the green dashed curve. **f.** Boosting wavelength-dependent compositional resolution, whose good agreement with the green dashed curve in **e** indicates the important role of light absorptivity for boosting the bonding force. **g.** Topography of a heteromaterial composed of h-BN and $WS_2$ multilayer nanoflakes from LvFM $f_1$ channel. **h.** Absorbance spectra of the multilayer h-BN and $WS_2$. **i,** The LvF amplitude mapping result of the h-BN and $WS_2$ mixture under 633 nm excitation with approximately 1.2 mW, where the white dash lines indicate the boundary between the h-BN and $WS_2$ areas.

**Summary**

Most versions of AFM developed so far work by manipulating tip-sample interactions or cantilever oscillations, apparently lacking of contributions from material performance reflected by bonding force, an even more important type of atomic force. We propose here the concept of LvF, as well as LvFM, which manipulates bonding forces by light boosting local atomic thermal motion beneath an AFM tip. As a result, the tip-sample vdW force is enhanced, and the enhancement factor is sensitively dependent on the atomic species composing the specimen, allowing for high compositional resolution. This makes LvFM particularly attractive for at least three facts: (i) large composition resolving power represented by a 20 dB enhancement factor and 150 times of the detection limit; (ii) high lateral spatial resolution down to sub-10 nm, which is even smaller than the tip size of 20 nm; (iii) low boosting laser power requirement and effectiveness in wide spectral range. These unique features exhibit that LvFM may be widely used for characterizing nano-materials, particularly 2D semiconductors, which are potential new generation of electronic and optoelectronic chip materials.

## Methods

### Specifications of the LvFM system

A tunable laser (Topaz-fps-50, BWT) incorporated with an acousto-optic tunable filter (AOTF) is employed as the light source. The output pulse width is 100 ps, which is modulated by a square wave at a chopping frequency $f_m$ by an acousto-optic modulator (AOM, AOMO 3080-125, Gooch & Housego). An aperture is mounted to select the 0th order beam of the AOM, as its propagation direction is independent of the wavelength. A linear polarizer (LP) and a half-wave plate (λ/2 WP) are used to adjust the polarization state of the laser beam. After passing through a beam splitter (BS), the laser beam is focused onto the sample surface from the bottom through an objective (Nikon, CF Plan 100x, NA = 0.95). A precision stage (P-517.3CL, Physik Instrumente) holding the sample moves laterally with a minimum step of 0.3 nm to realize the *x-y* scan. The mechanical vibration of a probe (PPP-NCHAu, Nanosensor or NSG10, TipsNano) is monitored by an AFM (NTEGRA, NT-MDT) with an optical lever. A modulation-demodulation technique is applied to extract the light boosting tip-sample force signal from the probe response, where two lock-in amplifiers (HF2LI, Zurich Instruments) are used to probe the first and second-order eigenmode vibration signals of the cantilever. A LED and a CCD are used for sample illumination and microscopic imaging, respectively, both of which are integrated inside an inverted microscope (Olympus, IX81).

### Preparation of the $MoTe_2$ sample

A sample of multi-layer $MoTe_2$ nanoflakes featuring in-plane 2H-1T' $MoTe_2$ homojunctions was fabricated utilizing Mo nanofilm as precursor via chemical vapor deposition (CVD). To initiate the process, Mo nanofilms were deposited onto a fused silica substrate by e-beam evaporation. The substrate, positioned face-down, was then carefully situated within the heating zone at the core of an alumina boat containing Te slugs (99.999%, Sigma Aldrich). Carrier gases comprising argon (99.999%) and hydrogen (99.999%) were employed during the reaction, with both gases flowing at a constant rate of 20 sccm (standard cubic centimeters per minute). The furnace was gradually heated to 600 °C over a period of 15 minutes (with a linear ramp rate of approximately 38.3 °C per min) and maintained at this temperature for 30 minutes to enable synthesis of the desired in-plane 2H-1T' $MoTe_2$ homojunctions. Subsequently, 10 minutes after the reaction, the furnace lid was opened to expedite the cooling process[36], allowing it to reach room temperature rapidly. X-ray diffraction (XRD, smartlab, Rigaku) characterization was performed, as shown in Fig. S5 in Supporting Information, confirming the existence of 1T' phase and 2H phase $MoTe_2$.

## Acknowledgements

This work is supported by the National Natural Science Foundation of China (62175121 and 61960206003), and by Tsinghua- Foshan Innovation Special Fund (2021THFS0102).

# Supporting Information

# Light-enhanced van der Waals force microscopy

Yu-Xiao Han, Benfeng Bai*, Jian-Yu Zhang, Jia-Tai Huang, Peng-Yi Feng, Hong-Bo Sun*

*State Key Laboratory of Precision Measurement Technology and Instruments, Department of Precision Instrument, Tsinghua University, Beijing 100084, China.*

*Corresponding author: baibenfeng@tsinghua.edu.cn; hbsun@tsinghua.edu.cn.*

## 1. Difference of LvFM with other methods

Different from AFM and other scanning probe microscopy methods such as scanning near-field optical microscopy (SNOM)[1, 2], the light-enhanced van der Waals force microscopy (LvFM) can be regarded as an optical version AFM that uses a tapping-mode AFM probe working in the tip-sample van der Waals (vdW) force and bonding force interaction regime to probe the subtle tip-sample force change caused by local molecular thermal fluctuation of materials under light illumination. Since different materials have different optothermal responses, the detected force contrast in different regions of a sample can reflect the compositional information of it. Based on this principle, the composing materials of a sample can be discriminated provided that the detected tip-sample force contrast is distinguishable.

To characterize heterostructures of nanomaterials, some high-resolution microscopic methods such as scanning electron microscopy (SEM) and scanning tunneling microscopy (STM) have been used to characterize nanoscale regions that may lead to an increase or decrease of the electric conductivity of 2D materials[3, 4]. However, despite the high spatial resolution of these methods for topography mapping, they have to be implemented in vacuum and cannot provide information about the associated electronic, spin, and optical responses of 2D materials.

AFM, though powerful in profiling surface features of samples with very high resolution, cannot probe the compositional information of materials. In the past decades, AFM has been combined with other techniques to study the dynamic, thermodynamic, optical, electric, magnetic, and chemical properties of materials[5]. For example, in order to characterize the chemical composition of materials, a technique derived from AFM has been developed, called the AFM-based infrared spectroscopy (AFM-IR)[6], which integrates the chemical analysis capability of infrared spectroscopy and the high spatial resolution of AFM. AFM-IR uses an AFM tip to probe the local thermal expansion of material caused by molecular vibrational absorption of infrared light[7]. Since molecular vibrational absorption usually occurs within the conduction band or valence band, it is associated with an intraband transition. Under infrared light excitation, the thermal expansion of sample surface may trigger the mechanical vibration of the probe usually working in contact mode on sample surface, from which the probe vibration response is analyzed to obtain the composition of the sample. As per the AFM-IR principle, the optothermal expansion of materials, such as polymers[6], biological tissues[7], and organics[8], should be large enough to be detectable. For this reason, the excitation light wavelength should be chosen to coincide with the molecular absorption energy transition levels (usually in infrared band) so as to produce a series of narrow fingerprint peaks with sufficiently large imaging contrast and signal-to-noise ratio (SNR). Nevertheless, due to the limitation of excitation laser power and the damage threshold of some materials, many samples cannot produce sufficiently large thermal expansion for AFM-IR. These restrict the application scenarios of AFM-IR.

Unlike AFM-IR, LvFM can respond to extremely small fluctuation of molecular thermal motions based on its vdW force sensing mechanism. Therefore, the light absorption in materials through non-radiative interband transition can be utilized, although it is usually much weaker than the molecular vibrational absorption in AFM-IR. Different from intraband absorption, the interband absorption is non-resonant, which produces a wide absorption band covering the visible and infrared regime. This feature greatly facilitates the choice of excitation light source in LvFM.

In addition to the above methods, there are also some other optical techniques such as tip-enhanced photoluminescence (TEPL)[9] and tip-enhanced Raman scattering (TERS)[10] that can be used to characterize the chemical composition of materials. However, these methods have to rely on both the emission (photoluminescence or Raman) properties of materials and spectroscopic scanning. In this regard, LvFM does not need to collect any optical signal from the sample (except for the mechanical vibration of the cantilever monitored by the optical lever) and does not need to use spectrometer as well. It is thus a very simple and versatile method for discriminating a wide range of materials including metals, dielectrics, semiconductors, and various nanomaterials such as 2D materials.

In recent years a near-field technique called photo-induced force microscopy (PiFM) has been proposed, which combines the advantages of AFM and near-field spectroscopy to perform near-field characterization with high spatial and spectral resolutions[11]. The principle of PiFM is that it detects the mechanical force generated between a photo-induced dipole at the sample surface and a mirror dipole in the AFM tip[12]. The mechanical vibration of the cantilever is driven by a dither piezo and modulated by a laser beam irradiating the tip-sample gap. The AFM probe works in non-contact mode, so that the net force between tip and sample is attractive. Different from the vdW force sensing mechanism of LvFM, the dipole-dipole interaction forces in PiFM are dominant only in materials with large polarizability and are also sensitive to the wavelength of illumination light.

With the above comparisons and discussions, it is seen that the LvFM is a brand-new non-radiative, non-destructive, and non-spectroscopic near-field method for super-resolution imaging of heterogenous composition of materials through near-field vdW force detection.

## 2. Analysis of interaction force on the tip in tapping mode AFM

In AFM, many forces play significant roles in the tip-sample interaction process. As shown in left part in the Fig. S1., the dominating interaction forces can be attributed to different categories according to the distance between the probe tip and sample surface, among which the chemical interaction force, the vdW force, and the electrostatic force are three commonly existing forces playing main roles in AFM[5]. The chemical interaction is a short-range repulsive force caused by Pauli repulsion. The vdW force and electrostatic force take effect in longer range and stay attractive as the tip approaches the sample surface, among which the interaction range of electrostatic force is wider and the vdW force commonly exists in various materials.

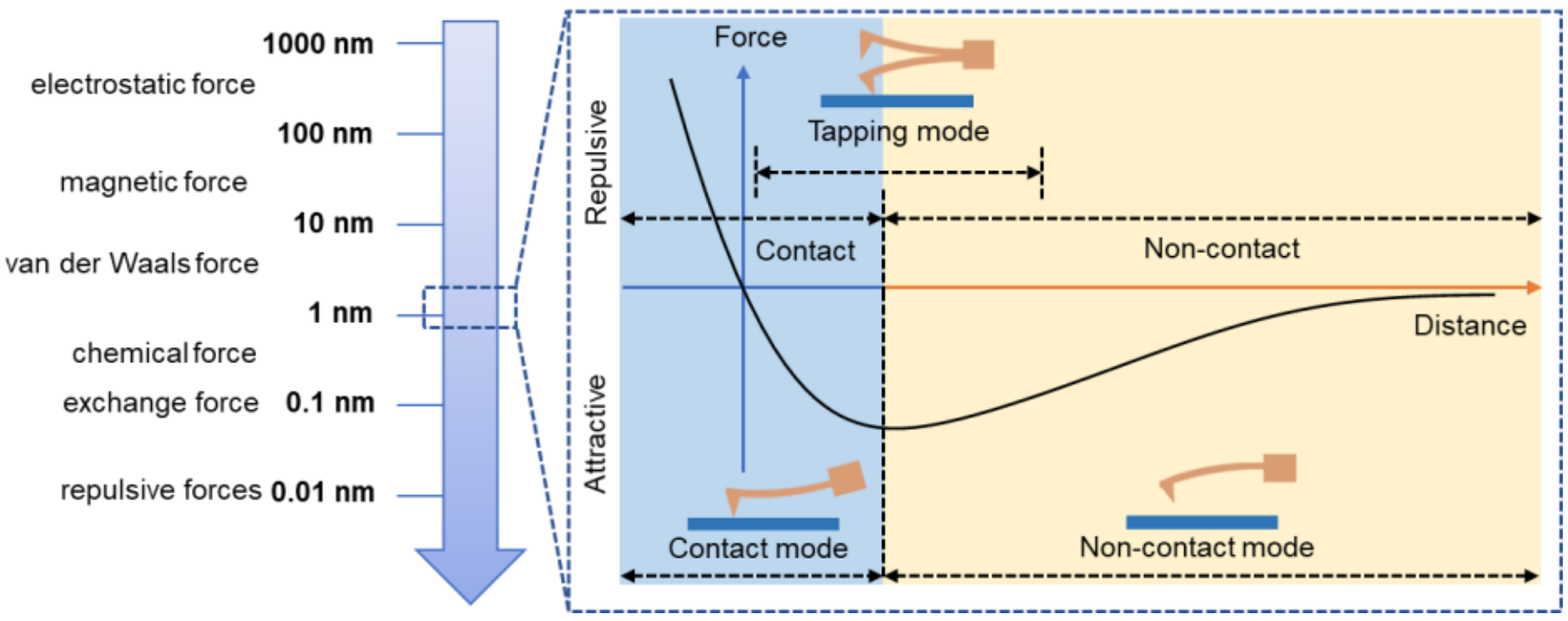

Fig. S1. Various distance-dependent tip-sample interaction forces. Left panel: various long- and short-range interaction forces between the probe tip and sample; Right panel: the distance-dependent interaction force curve of an AFM probe where three working modes are indicated.

The right part of Fig. S1 depicts the interaction force curve versus tip-sample distance at the region where the vdW force and partly chemical force are dominant. At different distances, the total force may be attractive or repulsive, depending on whether the attractive vdW force or the repulsive contact force are dominant. Correspondingly, the force curve can be divided into two regions, the non-contact region the and contact region, according to whether the gradient of the force curve is positive or negative, respectively, as indicated in the right panel of Fig. S1. Usually, the AFM probe works in tapping mode, which reciprocates in the non-contact and contact regions and the attractive and repulsive forces act alternately on the probe tip.

## 3. Numerical calculation of tip-sample force between the tip and sample

We established a model to describe and analyze the interaction force between an AFM tip and a sample when they are very close to each other. Detailed calculations have been reported in previous literatures[13, 14]. The force between two closely spaced molecules can be expressed as the Eq. (1) in the main text. Due to the additivity principle of vdW force, the total force between the tip and sample can be calculated from the combined force produced by all molecules of tip and sample, shown as the Eq.(2) in the main text. Schematic diagram of the calculation is shown as Fig. S2, with the sample is simplified as a flat plate of thickness $d$. The tip apex is seen as a hemisphere with the radium of $R$. The position coordinate of any atom (such as A point in the Fig .S2) in the tip is written as A ($\cos\alpha \cos\beta\, r_t$, $\cos\alpha \sin\beta\, r_t$, $\sin\alpha\, r_t$). The position coordinate of any atom (such as B point in the Fig .S2) in the sample is written as B ($\cos\zeta\, \rho$, $\sin\zeta\, \rho$, $z$). So the distance between the two atoms can be expressed as

$$r = \sqrt{(\cos\alpha\cos\beta r_t - \cos\zeta\rho)^2 + (\cos\alpha\sin\beta r_t - \sin\zeta\rho)^2 + (sin\alpha r_t - z)^2}\ , \qquad \text{(S1)}$$

where $\alpha$ and $\beta$ represent the angles of the x-axis and z-axis along the line from point A to the origin respectively, $\zeta$ represent the angle in the coordinate system of column. So the Lennard-Jones force between the atom A and atom B can be calculated by the Eq. (1) in the main text. The total force between tip and sample is

$$F = \int_h^{h+d}\int_0^{2\pi}\int_0^{+\infty}\left(\int_0^{\pi/2}\int_0^{2\pi}\int_0^{R} n_2 f_{\mathrm{LJ}}(r(\alpha,\beta,r_t,\zeta,\rho,z))dr_t d\alpha d\beta\right)d\rho d\zeta dz\ , \qquad \text{(S2)}$$

where $n_1$ ($n_2$) represent the molecule density and integral volume of the probe tip (the sample), respectively, $f_{\mathrm{LJ}}$ here refers to the force between a molecule in the probe tip and a molecule in the sample. $h$ is the distance between the tip apex and sample surface.

The calculated interaction force when $h$ ranges from 0.3 nm to 15 nm is shown as the blue curve in Fig. 1b. The calculation was performed with $\varepsilon = 1\times10^{-19}$ J and $r_0 = 0.3$ nm in Eq. (1), which are estimated values applicable to most common materials. The calculation results obtained are of general applicability. The calculation was performed in two dimensions, the probe tip is approximated as a semicircle with a radius of 10 nm. The step size (mesh precision) used in the calculation is 0.5 nm, and the software used is Matlab R2020b.

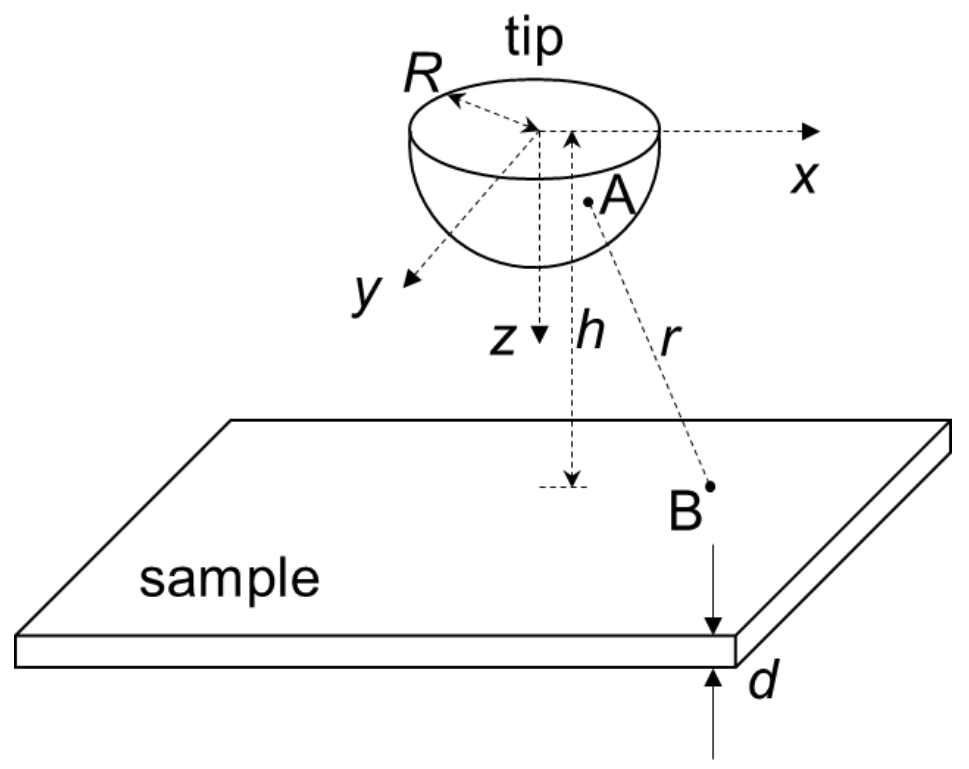

Fig. S2. Models for calculating vdW forces between the tip and the sample. The proportions in the figure do not represent the real situation.

## 4. Calculation of the temperature in the sample

Under light illumination, the thermal motions of molecules on a sample surface are intensified, which leads to local temperature rise in the illuminated region of the sample. The temperature rise can be modeled by the heat diffusion equation[15] with a heat source $Q(t)$:

$$\rho C\frac{dT}{dt} = \frac{Q(t)}{V(t)} - \kappa_{\text{eff}}\nabla^2 T(\rho, z, t)\,, \tag{S3}$$

where $\rho$, $C$, $T(\rho,z,t)$, and $\kappa_{\text{eff}}$ are the density, heat capacity, Kelvin temperature, and effective thermal conductivity[16] of the sample material, respectively, and $V(t) = V_p(t) + V_a(t) + V_d(t)$ is the volume of the sample material affected by the optothermal effect induced by light illumination, as shown in note 7 in this supporting information. In LvFM, the illuminating laser is modulated at frequency $f_m$, which can be described by rectangular pulses of duration $\tau_{\text{pulse}}$. Then the heating source can be expressed as

$$Q(t) = P_{\text{abs}} M(\tau_{\text{pulse}})\,, \tag{S4}$$

where $P_{\text{abs}}$ is the absorbed light power and $M(\tau_{\text{pulse}})$ is a rectangular box function of length $\tau_{\text{pulse}}$[15]. For most materials, the absorption of light is determined by the imaginary part of the complex permittivity. Then the absorbed light power can be expressed by

$$P_{\text{abs}} = \int \frac{1}{2}\omega|E|^2\,\text{Im}(\varepsilon)dV\,, \tag{S5}$$

where $\omega$, $|E|$, and $\text{Im}(\varepsilon)$ are the light frequency, the electric field amplitude on the sample surface, and the imaginary part of the material complex permittivity, respectively. The integral covers the entire volume illuminated by the light.

Rigorous modeling of this optothermal process requires complete multi-dimensional solution of Eq. (S1) to get the temperature distribution in the sample. By ignoring the heat accumulation and heat dissipation in substrate and in air around the sample, the model can be simplified so that the temperature under the probe apex can be estimated as[16]

$$T(0,z,t)=T_0+\Delta T_{\max}\frac{1-\mathrm{e}^{-t/\tau_{\mathrm{rel}}}}{1-\mathrm{e}^{-\tau_{\mathrm{pulse}}/\tau_{\mathrm{rel}}}}\cos(\zeta z/d) \qquad (t<\tau_{\mathrm{pulse}}), \qquad \text{(S6)}$$

$$T(0,z,t)=T_0+\Delta T_{\max}\mathrm{e}^{-(t-\tau_{\mathrm{pulse}})/\tau_{\mathrm{rel}}}\cos(\zeta z/d) \qquad (t>\tau_{\mathrm{pulse}}), \qquad \text{(S7)}$$

where $T_0$, $\Delta T_{\max}$, $\tau_{\mathrm{rel}}$, $\zeta$, and $d$ are the ambient temperature, the maximum temperature variation, the thermal relaxation (cooling) time, the thermal mode shape, and the thickness of the sample, respectively. $\Delta T_{\max}$ can be estimated as $\Delta T_{\max} \approx \tau_{\mathrm{rel}}P_{\mathrm{abs}}/\rho CV$ for $\tau_{\mathrm{rel}} < \tau_{\mathrm{pulse}}$ and $\Delta T_{\max} \approx \tau_{\mathrm{pulse}}P_{\mathrm{abs}}/\rho CV$ for $\tau_{\mathrm{rel}} > \tau_{\mathrm{pulse}}$. $\tau_{\mathrm{rel}}$ represents the time of getting heat equilibrium between the sample and the environment in the thermal diffusion process. In the absence of inter-facial thermal resistance, $\tau_{\mathrm{rel}}$ is given by the following equation[16]

$$\tau_{\mathrm{rel}} \approx \frac{4}{\pi^2}\frac{C\rho d^2}{\kappa_{\mathrm{eff}}}. \qquad \text{(S8)}$$

The temperature rise is a manifestation of the intensified motions of sample molecules under light illumination.

## 5. Theoretical model of the light-enhanced van der Waals force

The vdW force and light-enhanced van der Waals force (LvF) between a probe tip and a sample, when they are close to each other, are modeled by Eqs. (1)-(3) in the main text. The detailed derivation of the LvF is given below.

As discussed in the main text, the optothermal effect leads to an overall variation of the tip-sample distance $\Delta h$. In the condition of thin sample thickness ($d < \lambda$) and low power laser irradiation, $\Delta h$ is proportional to the temperature rise $\Delta T$ under the tip apex, which is described as

$$\Delta h=\mu d\Delta T=\mu d[T(0,z,t)-T_0], \qquad \text{(S9)}$$

where $\mu$ is the thermal expansion coefficient of material.

Since the AFM probe works in tapping mode, when the probe is very close to the sample surface, the tip-sample interaction force alternates between the attractive force and the repulsive force during an oscillation period, according to the curve force in Fig. 1b in the main text. The exemplar attractive force is the vdW force $F_{\mathrm{V}}$ and the exemplar repulsive force is the contact force $F_{\mathrm{C}}$ that can be described by the Derjaguin-Muller-Toporov (DMT)[14] model. Then the total tip-sample interaction force can be written as

$$F(z)=F_{\mathrm{V}}(z)+F_{\mathrm{C}}(z). \qquad \text{(S10)}$$

Specifically, $F(z)$ is derived as[14, 17]

$$F(z)\approx-\frac{H_{\mathrm{eff}}R}{12}\frac{1}{z^2} \qquad (z\ge r_0), \qquad \text{(S11)}$$

$$F(z)\approx-\frac{H_{\mathrm{eff}}R}{12}\frac{1}{r_0^2}+\frac{4}{3}Y^*\sqrt{R}\eta^{\frac{3}{2}} \qquad (\delta=r_0-z>0), \qquad \text{(S12)}$$

where $r_0$, $R$, $\eta$, $H_{\mathrm{eff}}$, and $Y^*$ are the position of potential minimum (~0.3 nm) which depends on material[18], the radius of probe tip apex, the indentation, the effective Hamaker constant of the sample material, and the effective Young's modulus of the sample material, respectively. Note that although Eqs. (S11) and (S12) are different from Eq. (2) in the main text, they all describe the tip-sample interaction force.

Using taylor formula, the tip-sample interaction force variation (LvF) induced by light boosting

is expressed as

$$\Delta F(z) \approx \frac{\partial F(z)}{\partial z}\Delta h \,. \tag{S13}$$

It shows that the LvF is proportional to $\Delta h$ and is a function of the tip-sample gap distance. According to the previous discussions, $\Delta h$ is closely related to the light absorption which is determined by the imaginary part of the dielectric constant of material. Therefore, LvF can be used for non-radiative optothermal imaging.

## 6. Mechanical resonances of the cantilever

A dual-modal AFM cantilever is used in the LvFM setup. The mechanical resonance frequencies of the first-order and second-order eigenmodes of the cantilever are denoted as $f_1$ and $f_2$, respectively. The tapping frequency of the probe tip is set as $f_1$ = 235.7 kHz, the demodulation frequency of the LvF signal is set as $f_2$ = 1471.8 kHz, and the laser modulation frequency is set as $f_m = f_2 - f_1$ = 1236.1 kHz. The oscillation amplitude of the probe at $f_2$ can be obtained as[17]

$$A_2 = \frac{d\Delta F(z)}{dz}\frac{1}{2\sqrt{m^2(\widehat{\omega}_2^2 - \omega_2^2)^2 + \widehat{b}_2^2\omega_2^2}}A_1 \,, \tag{S14}$$

where $m$ is the mass of the cantilever, $\omega_2$ is the effective angular frequency of the second eigenmode, $\widehat{\omega}_2$ is the effective angular frequency at $f_2$ under external force, $\widehat{b}_2$ is the effective damping coefficient of the cantilever at $f_2$ under the external force, and $A_1$ is the carrier amplitude. Since $f_2$ is the demodulation frequency, the second-order resonance amplitude $A_2$ of the cantilever is retrieved to represent the LvF signal.

## 7. Further explanation of sideband demodulation

As shown in Fig. S3, the volume of sample material affected by the optothermal effect can be divided into three regions: the central region $V_p$ right below the probe tip where the sample has the largest optothermal effect and the strongest force interaction with the probe, the ordinary light absorption region $V_a$ where the material directly absorbs light but with negligible tip-sample interaction, and the thermal diffusion region $V_d$ where the material receives heat not directly from the illumination but through heat diffusion from the nearby regions with higher temperature. When the sample is illuminated by a modulated laser beam, all the three regions have thermal response with the same modulation frequency $f_m$. Meanwhile, since the probe tip vibrates at tapping frequency $f_1$, the response of the central region $V_p$ contains a frequency component $f_1$. $f_m$ and $f_1$ couple with each other, resulting in new vibration frequencies $|f_m \pm f_1|$. Therefore, in LvFM, we can use a dual-modal cantilever with two mechanical eigenmode resonance frequencies $f_1$ and $f_2$, as schematically shown in Fig. 2c in the main text. The modulation frequency $f_m$ is chosen such that it satisfies $f_2 = |f_m \pm f_1|$. Then, the LvF signal can be demodulated from the resonance amplitude of the cantilever at $f_2$ with the best SNR.

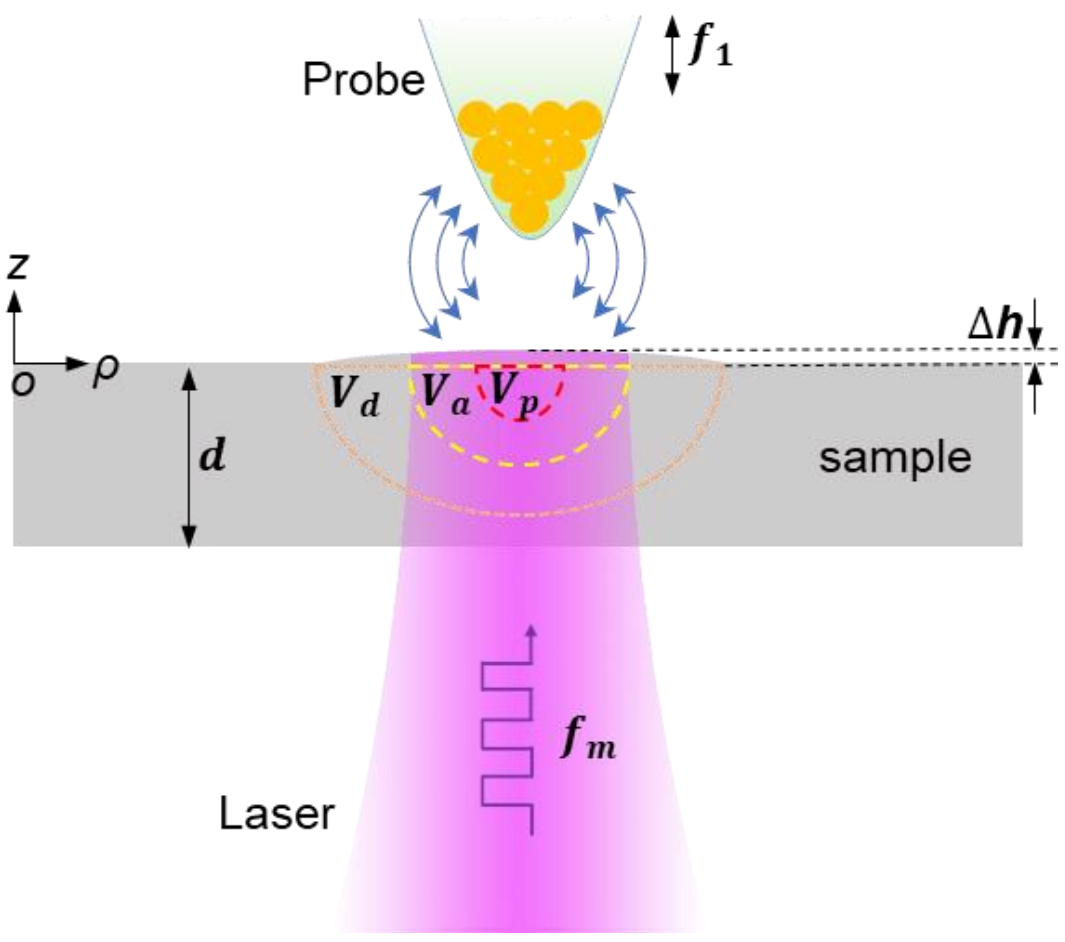

Fig. S3. Schematic of the tip-sample interaction in LvFM, where $V_p$ is the central region with the strongest light-boosting force response, $V_a$ is the region where the sample directly absorbs light but with negligible light-boosting force interaction with the probe, $V_d$ is the region where the sample receives heat through diffusion from the nearby regions, $f_1$ is the tapping frequency of the probe, $f_m$ is the modulation frequency of the illuminating laser, $d$ is the thickness of the sample, and $\Delta h$ is the overall variation of the tip-sample distance under laser irradiation.

## 8. Resolving power of the LvFM

In order to evaluate the resolving power of the LvFM, for the convenience of quantitative analysis, we can consider a specific 2D material sample, a $MoTe_2$ film composed of 2H phase and 1T' phase. The contribution to the total tip-sample force from each location $\rho$ on the sample, i.e., $F(\rho)/F$, is calculated numerically by the same procedure as stated above. The resultant force curves of 2H and 1T' phases are represented by the blue dashed curve and red dotted·curve in Fig. S4, respectively, each of which has a sharp peak right beneath the probe tip. Surprisingly, the full width at half maxima (FWHM) of each peak is only ~ 4.2 nm, which is even much smaller than the probe tip size (20 nm in diameter) and can be an estimate of the spatial resolution of LvFM. This reveals an important fact that, unlike AFM or SNOM whose spatial resolutions are mainly determined by the probe tip size, the LvFM has an extremely high spatial resolving power determined by the tip-sample vdW force interaction governed by Eq. 3 in the main text. It is for this reason that the spatial resolution of LvFM can be smaller than the probe tip size.

However, even though the spatial resolution is very high, the two phases of $MoTe_2$ cannot be distinguished in this case because the difference of the two force curves is negligibly small. Nevertheless, when the sample is illuminated by a light beam whose intensity profile is given by the black dashed curve in Fig. S4, the corresponding force curves of 2H and 1T' phases, i.e., $F'(\rho)/F'$, can be calculated and shown by the blue and red solid curves in Fig. S4, respectively. Now, on one hand, the peak intensities of both force curves are enhanced, validating the mechanism of LvF. On the other hand, their enhancement strengths are evidently different due to the different optothermal responses of the two phases, so that the two phases can be easily distinguished. Meanwhile, the resolving power of the method is retained. Actually, as can be seen from the FWHM values of the two force curves, which are 4.1 nm for the 1T' phase and 3.9 nm for the 2H phase, the spatial resolution is even increased a little.

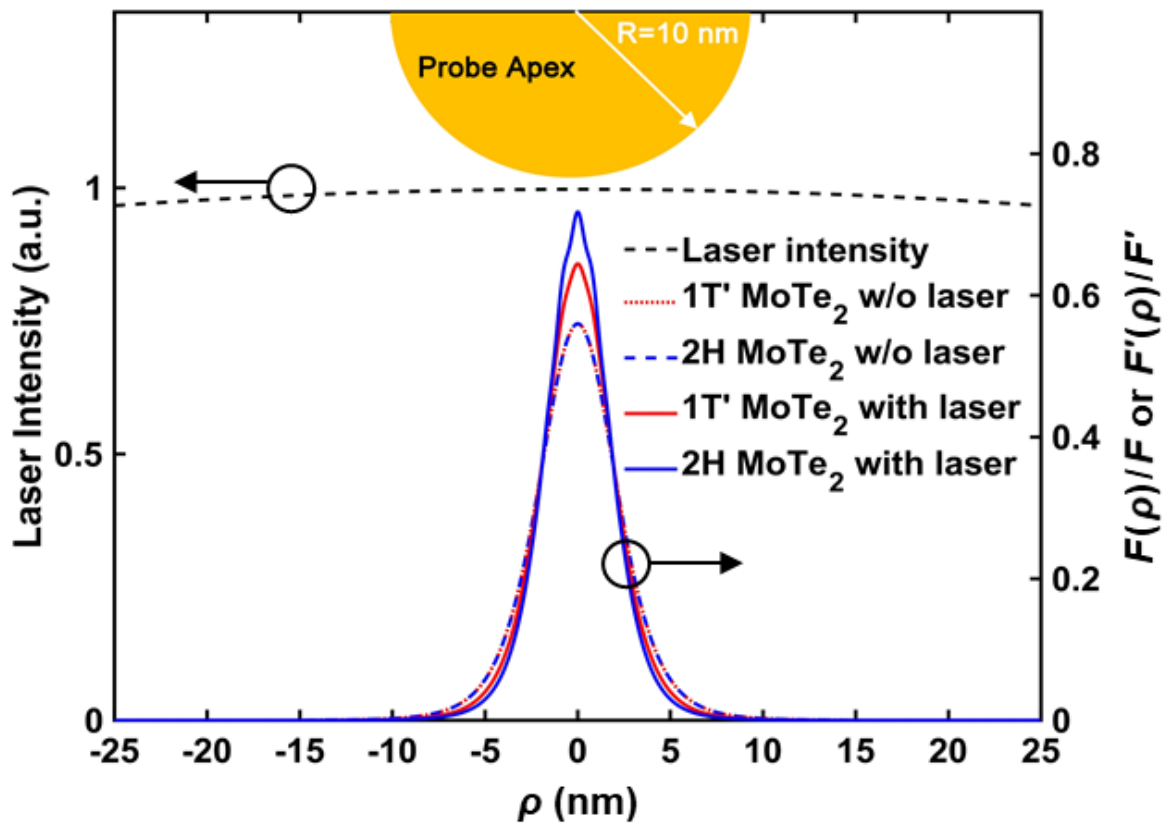

Fig. S4. Resolution power of the LvFM. Black dashed curve: intensity profile of the focused laser spot; Blue and red solid curves: calculated contribution to the total tip-sample force at each location $\rho$ of the sample surface; Blue dashed curve and red dotted curve: calculated contribution to the total force without laser participation at each location $\rho$ of the sample surface.

## 9. Characterisation of $MoTe_2$

X-ray diffraction (XRD, smartlab, Rigaku) characterization of the fabricated samples was performed, as shown in Fig. S5, which confirm the existence of 1T' phase and 2H phase $MoTe_2$ in the obtained samples. The XRD patterns in the Fig. S5 demonstrate that the samples are all polycrystalline monoclinic structures, with two diffraction peaks observed near $2\theta = 12.712°$ and 25.487° corresponding to the (002) and (004) diffraction surfaces of the polycrystalline cubic structure of $MoTe_2$, respectively [JCPDS data card #15-0658]. The XRD patterns of the films deposited in this study are in agreement with literature reports[19], confirming the formation of 1T' and 2H-$MoTe_2$ film. The films grew in (002) crystallographic plane selective orientation and showed good crystalline properties. Based on the XRD patterns we calculated the grain sizes of the 1T' and 2H-$MoTe_2$ films and the exact values are listed in Table S1.

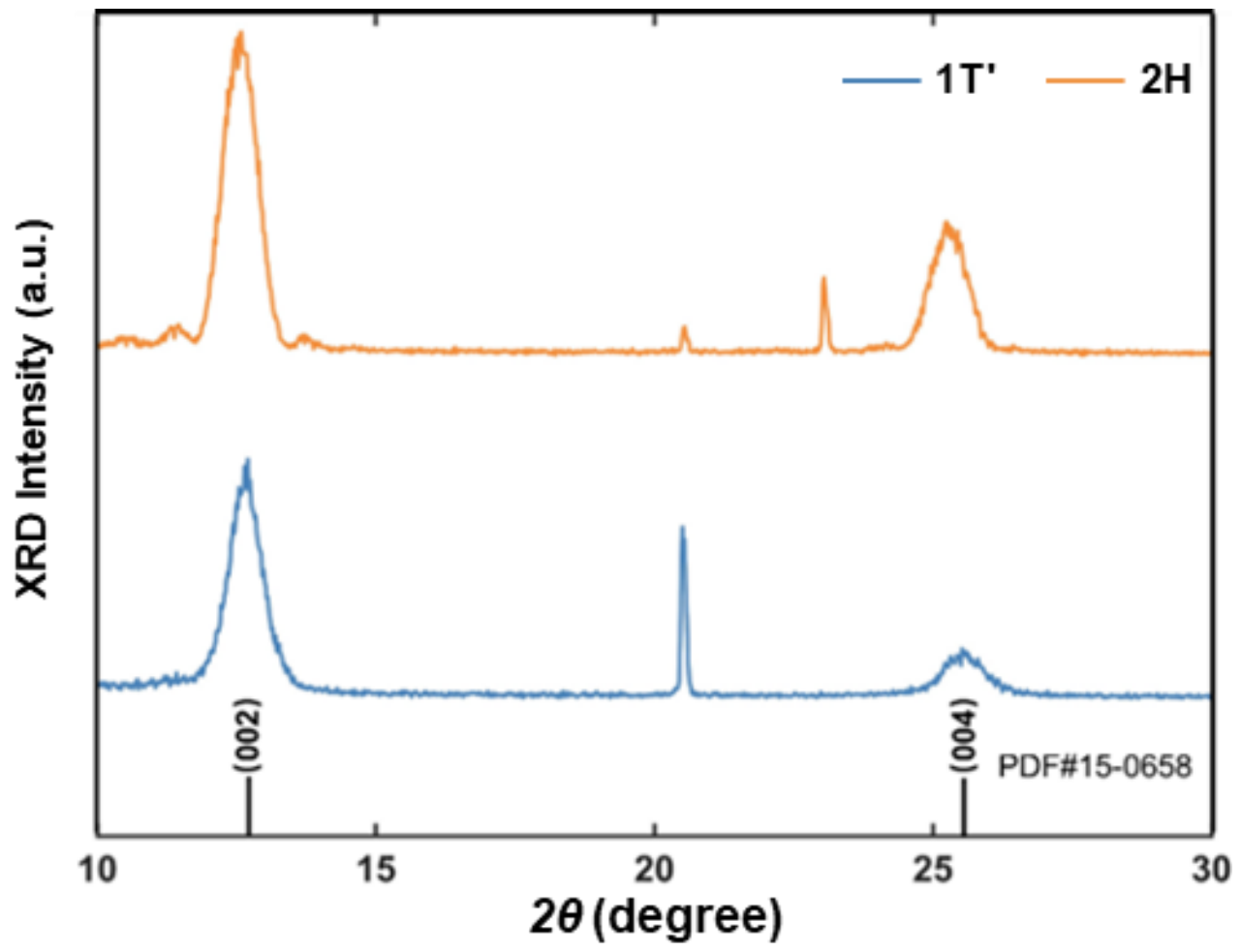

Fig. S5. XRD patterns of 1T' and 2H-$MoTe_2$ films grown on sapphire substrate.

| | $2\theta$ (°) | FWHM (°) | Grain size (nm) |
|---|---|---|---|
| 1T' | 12.712 | 0.696 | 11.357 |
| 2H | 12.781 | 0.702 | 11.261 |

Table. S1. Measured crystallographic parameters about 1T' and 2H-$MoTe_2$ films grown on sapphire substrate.

## 10. LvFM characterization of the $MoTe_2$ sample under different excitation wavelengths

As discussed in the main text, the LvFM is a broadband technique insensitive to the wavelength of excitation light. The composition materials of a heterostructure can be discriminated provided that the optothermal effects of the composing materials can provide sufficiently large LvF contrast. To demonstrate this fact, we have measured LvF images of the $MoTe_2$ sample under 9 different excitation wavelengths in the spectral range of 500 nm ~ 900 nm, as shown in Fig. S6. The LvF contrast of $U_H/U_T$ was also calculated and noted in the subfigures. These data are also the source data for drawing the LvF contrast curve in Fig. 4f in the main text.

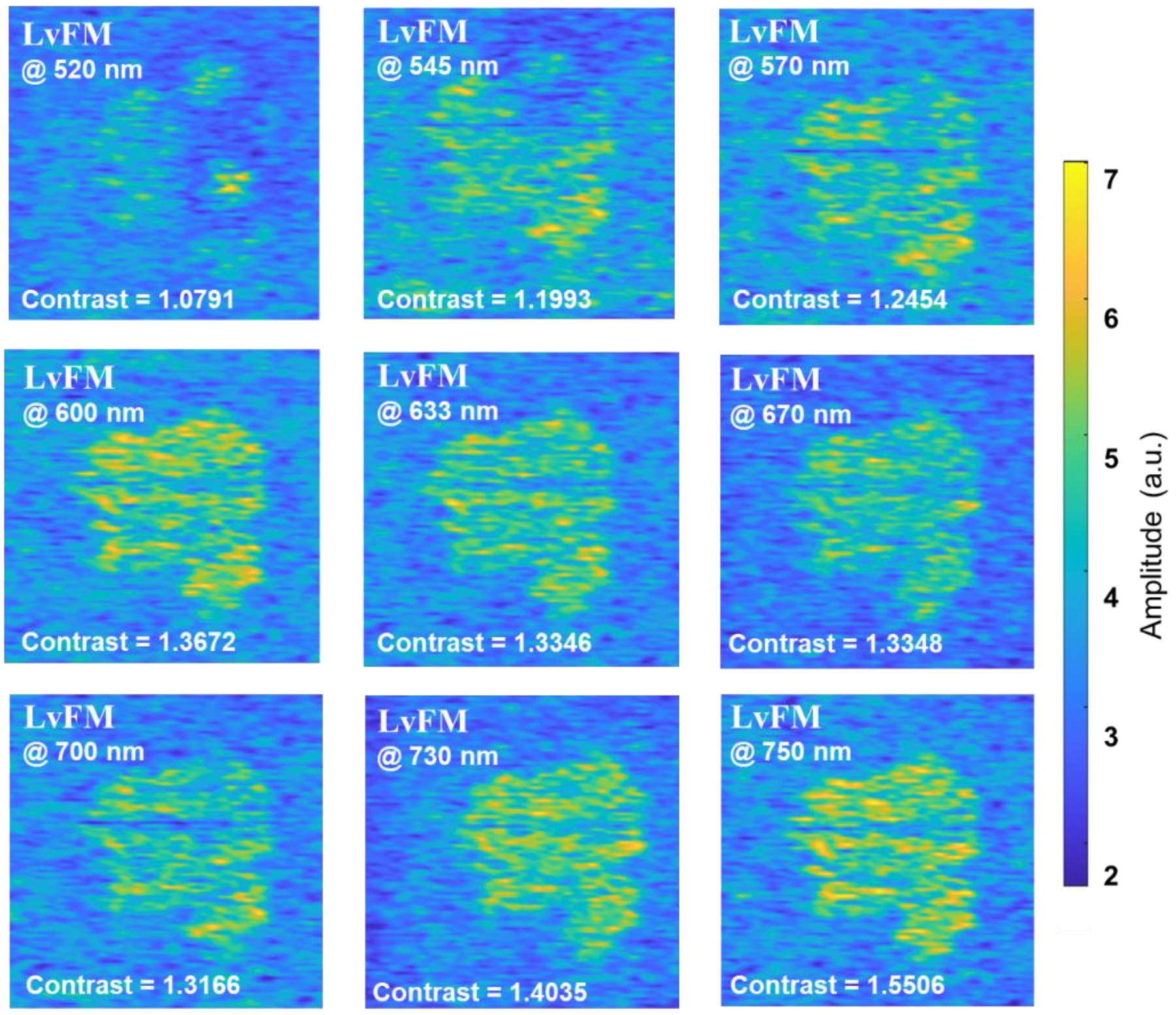

Fig. S6. LvF mapping images of the $MoTe_2$ film under 9 different excitation wavelengths.

## 11. LvFM characterization of the $WS_2$/h-BN heterogenous materials

LvF can be easily extended to the characterization of samples composed of heterogenous materials. Here, as a demonstration, we use LvFM to characterize a heterostructure of multilayer h-BN and $WS_2$ nanosheets. As is known, since h-BN has good chemical stability, good electrical insulation, and high thermal conductivity, it is often used as encapsulated layer, insulating layer, or substrate for TMD devices. The defects of h-BN encapsulated layer or insulating layer, even if in a small area, may cause unpredictable impact on TMD devices. Therefore, the super-resolution characterization of the heterostructure is highly demanded. In this experiment, a sample composed of h-BN and $WS_2$ multilayer nanoflakes were prepared on a borosilicate glass substrate. Borosilicate glass was used because it had a small thermal expansion coefficient and high light transmittance. As stated in the main text, $WS_2$ and h-BN have large light absorption coefficients difference in the wavelength from 600 ~ 680 nm. Therefore, LvF signal contrast should also be large in this band. This is verified by the LvF mapping results under excitation wavelengths of 600 nm and 680 nm, as shown in Fig. S7, respectively. Clearly, by the aid of LvF mapping, the two composing materials can be distinguished easily with very high spatial resolution down to about 10 nm.

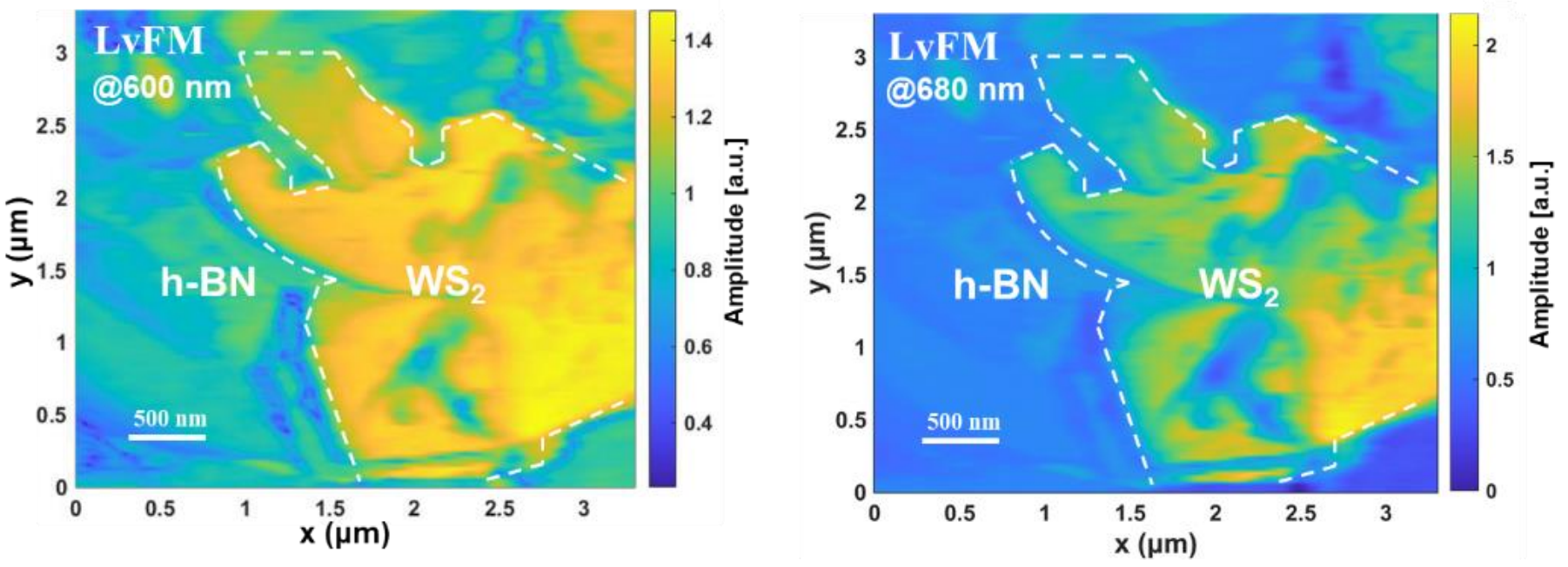

Fig. S7. LvF mapping images of the $WS_2$/h-BN heterostructure film at the wavelength of 600 and 680 nm.